\documentclass[aps,pra,preprint,groupedaddress]{revtex4-2}

\usepackage{graphicx,subfigure}
\newcommand{\eqref}[1]{(\ref{#1})}
\renewcommand{\varphi}{\wp} 
\usepackage{color}
\begin{document}


\title{Degenerate elliptic solutions of the quintic complex one-dimensional Ginzburg-Landau equation}


\author{Hans Werner Sch\"urmann}
\email[]{hwschuer@uos.de}
\affiliation{Department of Physics\\ University of Osnabr\"uck, Germany}

\author{Valery Serov}
\email[]{vserov@cc.oulu.fi}
\affiliation{Research Unit of Mathematical Sciences, University of Oulu, Finland\\
and Moscow Centre of Fundamental and Applied Mathematics -- MSU, Russia}



\begin{abstract}
A subset of traveling wave solutions of the quintic complex Ginzburg-Landau equation (QCGLE) is presented in compact form. The approach consists of the following parts.\\
 -- Reduction of the QCGLE to a system of two ordinary differential equations (ODEs) by a traveling wave ansatz.\\
 -- Solution of the system for two (ad hoc) cases relating phase and amplitude.\\
 -- Presentation of the solution for both cases in compact form.\\
 -- Presentation of constraints for bounded and for singular positive solutions by analysing the analytical properties of the solution by means of a phase diagram approach.\\
The results are exemplified numerically.
    
\end{abstract}


\maketitle


\section*{Keywords}

Ginzburg-Landau equation, Weierstrass' elliptic function, phase diagram

\section{Introduction}

As a partial differential equation (PDE) the quintic complex Ginzburg-Landau equation (QCGLE) is one of the most studied nonlinear equations in physics. Apart from many applications in the natural sciences [1], [2] the equation is interesting in and of itself: as a nonintegrable, nonlinear PDE, it admits a reduction to an autonomous ODE described in the simplest case by introducing the new variable $z=x-ct$, where $x$ is a space coordinate and $t$ is time (traveling wave reduction). While integrable PDEs are "easy" to solve by the Inverse Scattering Transform, for nonintegrable PDEs no such method is known to obtain solutions. 

Probably, due to this reason, several non-perturbative methods have been proposed to find some particular solutions of (partly) nonintegrable systems ("tanh-method", "exponential-method", "Riccati-method", "Jacobi expansion-method", ...., see also [3] and references therein). A comprehensive treatment of the QCGLE using Panlev$\acute e$ analysis is presented in [4]-[7]. In particular, an algorithm able to provide in closed form all those traveling wave solutions that are elliptic or degenerate elliptic has been applied to the QCGLE (results in [4], Eq.(52) and in [5], Section 6).

As a starting point for perturbation calculations or stability analysis exact traveling solutions of nonlinear 
equations such as (2) below are very useful, however rare. Remarkably, if certain  constraints for the parameters
are satisfied, solutions can be derived. In [8]-[10], special relations between phase and amplitude of the traveling wave are assumed
in the solution ansatz, leading to particular analytical solutions. Following this path, we propose a relatively simple (ad hoc) approach 
which allows us to find exact analytical solutions not known in the relevant literature.

The rest of the paper is organised as follows. In Section II we reduce the QCGLE to a system of two ordinary differential equations, specify it for two particular (ad hoc) relations between phase function and amplitude, and describe the procedure to obtain the ansatz parameters $c,\omega, d, b_0,b_1$ and constraints of these relations. An exact solution together with its dependence on the parameters of the QCGLE is presented in Section III. For elucidation examples are presented in Section IV. In Section V, the paper concludes with comments on articles having a certain contact to the present paper and with suggestions for further investigations.

\section{Reduction of QCGLE to first order nonlinear ODEs}   

We seek particular (traveling wave) solutions [2], [8], [9], [10], [11]
\begin{equation}
\psi(x,t)=f(x-ct)e^{i(\phi(x-ct)-\omega t)},\quad c\in \mathbf{R},\quad \omega \in \mathbf{R}, 
\end{equation}
where $f$ and $\phi$ are real-valued, of the QCGLE
$$
i\frac{\partial}{\partial t}\psi (x,t)+(c_3+ih_3)|\psi(x,t)|^2\psi(x,t)+(c_5+ih_5)|\psi(x,t)|^4\psi(x,t)+
$$
\begin{equation}
+(c_1-ih_1)\frac{\partial^2}{\partial x^2}\psi (x,t)-i\epsilon \psi(x,t)=0,
\end{equation}
with complex-valued function $\psi(x,t)$ and with real dimensionless constants $\epsilon, h_1,h_3,h_5,$\\$c_1,c_3,c_5$. As mentioned in the Introduction, Eq.(2) has a wide range of applications. Thus the meaning of variables and parameters may be quite different (see [1], [8] and references therein). Usually, $x$ denotes the scaled propagation distance and $t$ the scaled time. 

To find real $f,\phi, c, \omega$, together with corresponding assumptions and constraints for their existence, we insert ansatz (1) into Eq.(2), set $z=x-ct$, introduce $F(z)=f^2(z),\tau(z)=\phi'(z)$, and separate real and imaginary parts.  Hence we obtain the system of two ordinary differential equations for functions $F(z)$ and $\tau(z)$
$$
4\omega F^2(z)+4c_3F^3(z)+4c_5F^4(z)+4c\tau(z)F^2(z)-4c_1\tau^2(z)F^2(z)+4h_1\tau(z)F(z)F'(z)-
$$
\begin{equation}
-c_1(F'(z))^2+4h_1\tau'(z)F^2(z)+2c_1F(z)F''(z)=0,
\end{equation}
$$
-4\epsilon F^2(z)+4h_3F^3(z)+4h_5F^4(z)+4h_1\tau^2(z)F^2(z)-2cF(z)F'(z)+4c_1\tau(z)F(z)F'(z)+
$$
\begin{equation}
+h_1(F'(z))^2+4c_1\tau'(z)F^2(z)-2h_1F(z)F''(z)=0.
\end{equation}
As mentioned above, there is a general method in the literature [5], {\color{red}{[12]}} to find all elliptic solutions of the QCGLE. We note however that the explicit solution obtained by this method (see Eq.(52) in [4b]) exists at the price of 5 constrains among the coefficients of the system (3)-(4). Thus it is obvious to try for a different approach.
In particular, we seek for solutions $F(z)$ of the system (3)-(4) by assuming [8], [9], [10] two possibilities for $\tau(z)$
\begin{equation}
\tau(z)=d\frac{F'(z)}{F(z)},
\end{equation}
\begin{equation}
\tau(z)=b_0+b_1F(z),
\end{equation}
where real constants $d, b_0, b_1$ are to be determined. 

We consider the solutions $F(z)$ of the system (3)-(4), with assumptions for $\tau(z)$ from above, and determine parameters $c,\omega, d, b_0,b_1$ for both cases (5), (6) separately. 
In each part (for (5) and for (6)) we derive the solution $\psi(x, t)$ and study its properties (disregarding its stability).
We note that system (3)-(4) is equivalent to (13)-(14) in [5].

To simplify formulas we use the abbreviations 
$$
D_1=c_1^2+h_1^2,\quad D_2=c_3h_1+c_1h_3,\quad D_3=c_1c_3-h_1h_3,\quad D_4=c_5h_1+c_1h_5,\quad D_5=c_1c_5-h_1h_5
$$ 
in what follows.

First, we consider Eqs.(3), (4) subject to $\tau(z)=d\frac{F'(z)}{F(z)}$.
Equations (3), (4) read in this case 
$$
4\omega F^2(z)+4c_3F^3(z)+4c_5F^4(z)+4cdF(z)F'(z)-c_1(1+4d^2)(F'(z))^2+
$$
\begin{equation}
+(4h_1d+2c_1)F(z)F''(z)=0,
\end{equation}
$$
-4\epsilon F^2(z)+4h_3F^3(z)+4h_5F^4(z)-2cF(z)F'(z)+h_1(1+4d^2)(F'(z))^2+
$$
\begin{equation}
+(4c_1-2h_1)F(z)F''(z)=0.
\end{equation}
By eliminating $F''(z)$ from (7) and (8), we obtain for $F'(z)$
\begin{equation}
F'(z)=F(z)\left(\pm\sqrt{a+bF(z)+fF^2(z)}+\frac{cc_1}{2dD_1}\right),
\end{equation}
with 
\begin{equation}
a=\frac{c^2c_1^2}{4d^2D_1^2}+\frac{2\epsilon(c_1+2dh_1)+2\omega(2dc_1-h_1)}{d(1+4d^2)D_1},
\end{equation}
\begin{equation}
b=\frac{4dD_3-2D_2}{d(1+4d^2)D_1},
\end{equation}
\begin{equation}
f=\frac{4dD_5-2D_4}{d(1+4d^2)D_1}.
\end{equation}
Using (9), we get 
\begin{equation}
F''(z)=F'(z)\left(\pm\frac{2a+3bF(z)+4fF^2(z)}{2\sqrt{a+bF(z)+fF^2(z)}}+\frac{cc_1}{2dD_1}\right).
\end{equation}
Substituting (9) and (13) to the system (7)-(8), we obtain the following equivalent system only in terms of $F(z)$ and the parameters of QCGLE
$$
4acdD_1(2h_1d+c_1)^2+[c^2c_1(c_1^2+4d^2(c_1^2+2h_1^2)+4c_1h_1d)+4ad^2D_1^2(c_1-4c_1d^2+4h_1d)+16\omega d^2D_1^2]
$$
$$
\sqrt{a+bF(z)+fF^2(z)}+F(z)[2bcdD_1(8h_1^2d^2+10c_1h_1d+3c_1^2)+8d^2D_1^2(2c_3+b(c_1-2c_1d^2+3h_1d))
$$
$$
\sqrt{a+bF(z)+fF^2(z)}]+F^2(z)[8cdD_1f(c_1^2+3c_1h_1d+2h_1^2d^2)+
$$
\begin{equation}
+4d^2D_1^2(4c_5+f(3c_1-4c_1d^2+8h_1d))\sqrt{a+bF(z)+fF^2(z)}]=0
\end{equation}
and
$$
4acdD_1(4c_1h_1d^2+2(c_1^2-h_1^2)d-c_1h_1)+[c^2c_1(4c_1h_1d^2-4h_1^2d-c_1h_1)+4ad^2D_1^2(4h_1d^2+4c_1d-h_1)-
$$
$$
-16\epsilon d^2D_1^2]\sqrt{a+bF(z)+fF^2(z)}+F(z)[2bcdD_1(8c_1^2d^2+(c_1^2-4h_1^2)d-3c_1h_1)+
$$
$$
+8d^2D_1^2(2h_3+b(3c_1d-h_1+2h_1d^2))\sqrt{a+bF(z)+fF^2(z)}]+
$$
$$
+F^2(z)[8cdD_1f(2c_1h_1d+(2c_1^2-h_1^2)d-c_1h_1)+
$$
\begin{equation}
+4d^2D_1^2(4h_5+f(8c_1d+4h_1d^2-3h_1))\sqrt{a+bF(z)+fF^2(z)}]=0.
\end{equation}
In Eqs.(14)-(15) the coefficients in front of $F(z)$, $F^2(z)$, $\sqrt{a+bF(z)+fF^2(z)}$ as well as free coefficients must be equal to zero, 
leading to equations which imply $c=0$ necessarily. With $c=0$ the system (14)-(15) can be simplified to the following system of six equations:
\begin{equation}
4\omega=a(4c_1d^2-4h_1d-c_1),\quad 4\epsilon=a(4h_1d^2+4c_1d-h_1),
\end{equation}
\begin{equation}
b(2c_1d^2-3h_1d-c_1)=2c_3,\quad b(2h_1d^2+3c_1d-h_1)=-2h_3,
\end{equation}
\begin{equation}
f(4c_1d^2-8h_1d-3c_1)=4c_5,\quad f(4h_1d^2+8c_1d-3h_1)=-4h_5.
\end{equation}
Combining the first Eq.(16) and (10), we obtain
\begin{equation}
\omega=\frac{\epsilon(4c_1d^2-4h_1d-c_1)}{4h_1d^2+4c_1d-h_1}.
\end{equation}
Furthermore, Eqs.(16)-(18) are solved by
\begin{equation}
d=\frac{-3D_3\pm\sqrt{9D_3^2+8D_2^2}}{4D_2}\quad and\quad d=\frac{-2D_5\pm\sqrt{4D_5^2+3D_4^2}}{2D_4}.
\end{equation}
Consistency of $d'$s in the equations (20) leads to the constraints
\begin{equation}
\frac{-3D_3\pm\sqrt{9D_3^2+8D_2^2}}{4D_2}=\frac{-2D_5\pm\sqrt{4D_5^2+3D_4^2}}{2D_4},
\end{equation}
necessary for the existence of solutions $d$.

Thus, parameters $d$, $\omega$ and $a, b, f$ in Eq.(9) are expressed according to (19)-(20) and (10)-(12), respectively in terms of the 
parameters of the QCGLE (with constraint (21)).
The solution $F(z)$ of Eq.(9) is presented below.

Second, considering the case $\tau(z)=b_0+b_1F(z)$, 
and inserting $\tau(z)$ into the system (3)-(4), we get
$$
4(\omega+cb_0-c_1b_0^2)F^2(z)+4(c_3+cb_1-2c_1b_0b_1)F^3(z)+4(c_5-c_1b_1^2)F^4(z)-c_1(F'(z))^2+
$$
\begin{equation}
+4h_1(b_0F(z)+2b_1F^2(z))F'(z)+2c_1F(z)F''(z)=0
\end{equation}
and
$$
4(h_1b_0^2-\epsilon)F^2(z)+4(h_3+2h_1b_0b_1)F^3(z)+4(h_5+h_1b_1^2)F^4(z)+h_1(F'(z))^2+
$$
\begin{equation}
+\left((4c_1b_0-2c)F(z)+8c_1b_1F^2(z)\right)F'(z)-2h_1F(z)F''(z)=0.
\end{equation}
Eliminating $F''(z)$ from (22) and (23), $F'(z)$ can be derived as
\begin{equation}
F'(z)=\frac{\left(h_1(\omega+cb_0)-c_1\epsilon\right)F(z)+(D_2+ch_1b_1)F^2(z)+D_4F^3}{\frac{cc_1}{2}-b_0D_1-2b_1D_1F(z)}.
\end{equation}
Assuming for simplicity $b_0=\frac{cc_1}{2D_1}$, we get
\begin{equation}
F'(z)=-\frac{h_1(\omega+\frac{c^2c_1}{2D_1})-\epsilon c_1+(ch_1b_1+D_2)F(z)+D_4F^2(z)}{2b_1D_1}
\end{equation}
and hence
\begin{equation}
F''(z)=-\frac{(ch_1b_1+D_2+2D_4F(z))F'(z)}{2b_1D_1}.
\end{equation}
Substitution of (25) and (26) into system (22)-(23) and considering vanishing coefficients of powers of $F(z)$, leads to 
\begin{equation}
2D_1(h_1\omega-c_1\epsilon)+c^2c_1h_1=0,
\end{equation}
\begin{equation}
-16b_1^4c_1D_1^2+3c_1D_4^2+16c_1b_1^2D_1D_5=0,
\end{equation}
\begin{equation}
c_1D_2D_4+4b_1^2c_1D_1D_3=0,
\end{equation}
$$
-2cc_1h_1b_1D_1D_2+c_1(c^2c_1h_1D_4+D_1(D_2^2-2c_1\epsilon D_4+2h_1\omega D_4))-
$$
\begin{equation}
-b_1^2D_1(c^2c_1(4c_1^2-3h_1^2)+16D_1c_1(h_1\epsilon+c_1\omega))=0,
\end{equation}
where $c\ne 0, h_1\ne 0, b_1\ne 0$ have been assumed. Parameters $\omega, b_1$ and $c$ must satisfy system (27)-(30). Eq.(27) implies 
\begin{equation}
\omega=\frac{2c_1\epsilon D_1-c^2c_1h_1}{2h_1D_1}.
\end{equation}
Solutions of (28) and (29) are 
\begin{equation}
b_1^2=\frac{2D_5+\sqrt{4D_5^2+3D_4^2}}{4D_1}
\end{equation}
and
\begin{equation}
b_1^2=-\frac{D_4D_2}{4D_1D_3},
\end{equation}
respectively. Inserting $\omega$ into (30) and solving for $c$ we get
\begin{equation}
c=\frac{-h_1D_2\pm2\sqrt{D_1D_2^2+4\frac{\epsilon}{h_1} b_1^2D_1^2(4c_1^2+3h_1^2)}}{b_1(3h_1^2+4c_1^2)}.
\end{equation}
Consistency of (32) and (33) yields the constraint
\begin{equation}
D_4D_2^2+4D_2D_3D_5-3D_4D_3^2=0.
\end{equation}
With (31), (32) (or (33)), (34) all parameters in (25) are determined in terms of the parameters of the QCGLE, so that (25) can be solved for $F(z)$ subject to (35).

\section{Traveling wave solutions}  

The nonlinear first order ODEs (9) (with $c=0$) and (25) can be solved by standard methods yielding $F(z)$ as an inverse function 
of an elliptic integral, but not $F(z)$ explicitly. Thus, it is obvious to look for another possibility to find elliptic solutions of (9) and (25). With $F'$
according to (9), $c=0$ necessarily, and hence, taking into account (10)-(12) and (16)-(21), the solution of system (7)-(8) uniquely can be 
rewritten as
\begin{equation}
(F'(z))^2=\alpha F^4(z)+4\beta F^3(z)+6\gamma F^2(z)
\end{equation}
with 
\begin{equation}
\alpha=f,\quad \beta=\frac{b}{4},\quad \gamma=\frac{a}{6}.
\end{equation}
Thus, Eq.(9) (with $c=0$) and Eq.(36) are equivalent. 

Following the same line with $F'$ according to (25) (using (22) or (23), (27)-(35)) we obtain (36), and hence equivalence of (25) and (36).
The coefficients in (36) are given by 
\begin{equation}
\alpha=\frac{D_4^2}{4b_1^2D_1^2},\quad \beta=\frac{(ch_1b_1+D_2)D_4}{8b_1^2D_1^2},\quad \gamma=\frac{(ch_1b_1+D_2)^2}{24b_1^2D_1^2}.
\end{equation}
The solution of (36) is well known (see \cite{SS1}, Eq.(17)). With some algebra we get
\begin{equation}
F(z)=F_0\frac{2\sqrt{\alpha F_0^2+4\beta F_0+6\gamma}\wp'(z)+4\wp^2(z)+(8\gamma+4\beta F_0)\wp(z)-2\gamma\beta F_0-5\gamma^2}{4\wp^2(z)-4\wp(z)(\alpha F_0^2+2\beta F_0+\gamma)+4F_0^2(\beta^2-\alpha\gamma)+4\beta\gamma F_0+\gamma^2},
\end{equation}
where $\wp(z; g_2, g_3)$ denotes Weierstrass' elliptic function with invariants 
$g_2=3\gamma^2,\quad g_3=-\gamma^3$
and $F_0=F(0)$ as an integration constant is the intensity $F(z)$ at $z=0$. The invariants $g_2, g_3$ can be expressed
in terms of the ansatz parameters and the coefficients of the QCGLE. For the case (5) we obtain 
\begin{equation}
g_2=\frac{a^2}{12},\quad g_3=-\frac{a^3}{216}
\end{equation}
with $a$ given by Eq.(10). For case (6) we get 
\begin{equation}
g_2=3\left(\frac{(ch_1b_1+D_2)^2}{24b_1^2D_1^2}\right)^2,\quad g_3=- \left(\frac{(ch_1b_1+D_2)^2}{24b_1^2D_1^2}\right)^3,
\end{equation}
with $b_1^2$ according to Eqs.(32) or (33) has been used. 

Integration of Eqs.(5) and (6), using Eq.(39), yields the phase function $\phi(z)$. For case (5) we obtain
\begin{equation}
\phi(x)=d\cdot log\left(F_0\frac{2\sqrt{\alpha F_0^2+4\beta F_0+6\gamma}\wp'(x)+4\wp^2(x)+(8\gamma+4\beta F_0)\wp(x)-2\gamma\beta F_0-5\gamma^2}{4\wp^2(x)-4\wp(x)(\alpha F_0^2+2\beta F_0+\gamma)+4F_0^2(\beta^2-\alpha\gamma)+4\beta\gamma F_0+\gamma^2}\right),
\end{equation}
where $\alpha,\beta,\gamma, g_2, g_3$ given by (37) and (40), respectively and chirp parameter $d$ by (20).
Since $c=0$ in this case $F$ and $\phi$ are stationary. As is known \cite{AA}, subject to certain conditions, 
traveling wave solutions can be obtained from the stationary solutions applying a Galilean transformation (see \cite{AA}, Eq.(53))
$$
x'=x+vt,\quad t'=t,\quad \psi'(x',t')=\psi(x,t)e^{i(vx+\frac{v^2}{2}t)},\quad v=const,
$$
to solutions (39) and (42). -- We disregard this possibility to find solutions of the QCGLE. 

For case (6), integral  
$\phi(z)=\int(b_0+b_1F(z))\,dz$ with $F(z)$ according to (39), cannot be evaluated in closed form (in general).
With respect to the example presented below, for particular $F_0$, integration yields a closed form result. 
If $F_0=-\frac{\beta}{\alpha}, \alpha>0,\beta<0$ (see example in Section IV), solution $F(z)$ reads 
\begin{equation}
F(z)=\frac{\frac{\beta^2}{\alpha\sqrt{\alpha}}\wp'(z)-\frac{2\beta}{\alpha}\wp^2(z)-\frac{2\beta^3}{3\alpha^2}\wp(z)+\frac{4\beta^5}{9\alpha^3}}{2\wp^2(z)+\frac{2\beta^2}{3\alpha}\wp(z)-\frac{4\beta^4}{9\alpha^2}}
\end{equation}
and
\begin{equation}
\phi(z)=\left(b_0-\frac{\beta}{\alpha}b_1\right)z+\frac{b_1}{2\sqrt{\alpha}}\left(log\left(\wp(z)-\frac{\beta^2}{3\alpha}\right)-log\left(\wp(z)+\frac{2\beta^2}{3\alpha}\right)\right),
\end{equation}
where $\alpha,\beta, g_2, g_3$ are given by (37) ($\gamma=\frac{2\beta^2}{3\alpha}$), (40) and by $b_0=\frac{cc_1}{2D_1}, b_1^2=-\frac{D_2D_4}{4D_1D_3}$
with $c$ according to (34). 

It should be noted that in both cases (40) and (41) the discriminant of $\wp(z;, g_2, g_3)$ vanishes, so that 
$\wp(z)$ degenerates to trigonometric ($g_2>0, g_3>0$) or hyperbolic ($g_2>0, g_3<0$) functions, thus depending on the sign of $\gamma$
in (37) and (38) (see \cite{Abr}, 18.12). Hence, $F(z)$ according to (39) must be rewritten in two versions. In the following, we are preferring
to maintain form (39) since it is rather compact and independent on the sign of $\gamma$.

Summing up, solutions (1) of Eq.(2) can be derived if $\phi'(z)$ and $F(z), z=x-\omega t,$ are related by (5) or (6), with $F(z)$ given by
(39) subject to certain constraints and conditions for $d, c, b_0, b_1, \omega$. It is necessary to check consistency of the results
with the initial assumptions $\{f, \phi, c, \omega, d, b_0, b_1\}\subset \mathbf{R}$. This will be done in the following.

First, we note for case (5) that $\alpha, \beta, \gamma, g_2, g_3$ are real (see Eq.(10) with $c=0$). Chirp parameter $d$ is real
and hence $\omega$ since (20) is satisfied. Constraint (21) is necessary for unique existence of $d$. 

Second, for case (6),  $\alpha, \beta, \gamma, g_2, g_3$ are real (see Eqs.(37), (40)) if $c$ is real according to Eq.(34) and hence $\omega$.
Thus, real $c$ implies nonnegative radicand in Eq.(34). Constraint (35) is necessary for unique $b_1^2$. -- We emphasise that real 
$\alpha, \beta, \gamma, g_2, g_3$ are important for evaluation of Eqs.(38), (39).

Third, we note that real $g_2, g_3$ imply real $\wp(z;, g_2, g_3)$ and  $\wp'(z;, g_2, g_3)$ if $z$ is real (see \cite{Abr}, 18.5). Thus, $F(z)$
according to (39), is real, since $\alpha F_0^2+4\beta F_0+6\gamma \ge 0$ due to Eq.(38).

Due to the properties of $\wp(z; g_2, g_3)$ (poles and periods) and the dependence of the denominator on $\alpha, \beta, \gamma, F_0$, 
a singularity analysis of $F(z)$ w.r.t. $z$ on the basis of (39) is very difficult. Indeed, what can be stated is that $F(z)$ exhibits only poles w.r.t. $z$,
consistent with a Theorem by Conte and Ng [7]. To conclude, there is consistency between the results obtained and the initial assumptions
for $f, \phi, c, \omega, d, b_0, b_1$.

As is known \cite{DSch}, that Eq.(38) alone is suitable to study the nature of the solution $F(z; F_0, \alpha, \beta, \gamma)$ by considering
the graphs $\{(F')^2, F\}$ of Eq.(38) (denoted as "phase diagram" in the following). Types are depicted in Fig.1.

\cleardoublepage
\begin{figure}
\subfigure[]{
\includegraphics[scale=0.4]{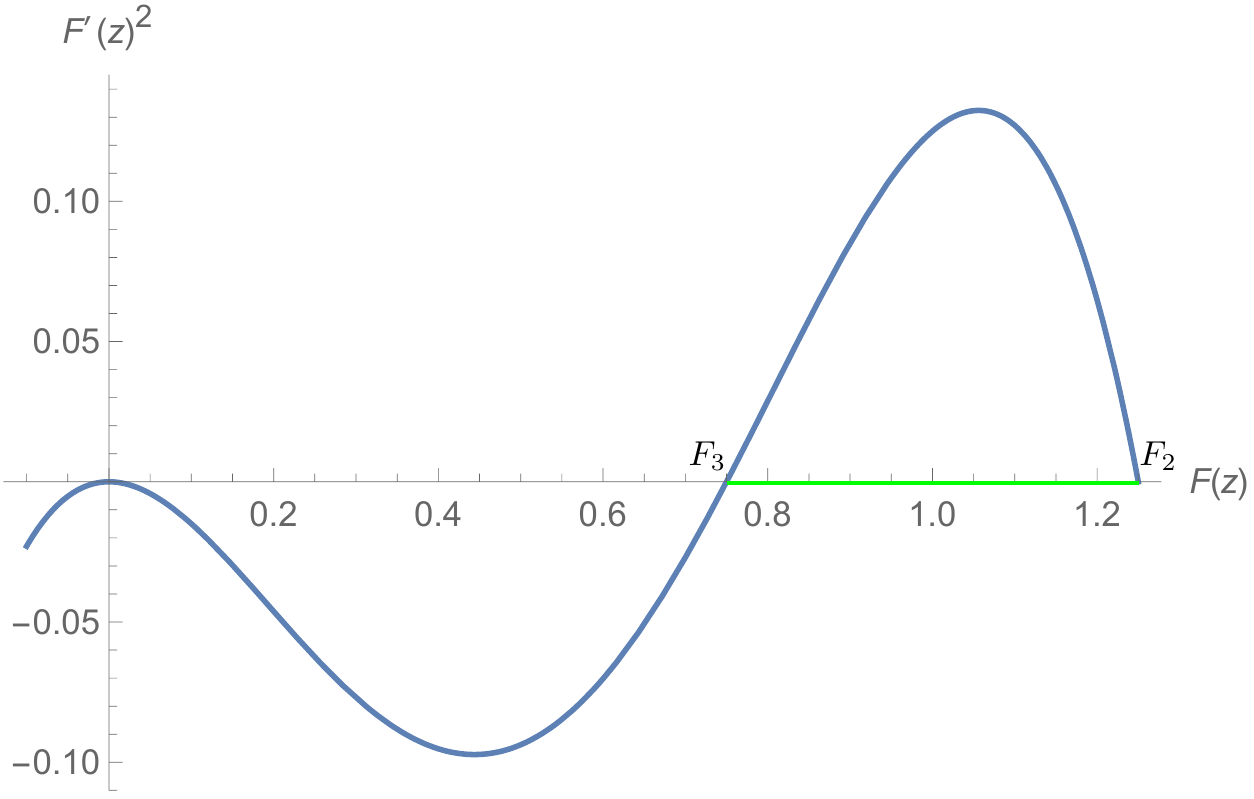}
}
\subfigure[]{
\includegraphics[scale=0.4]{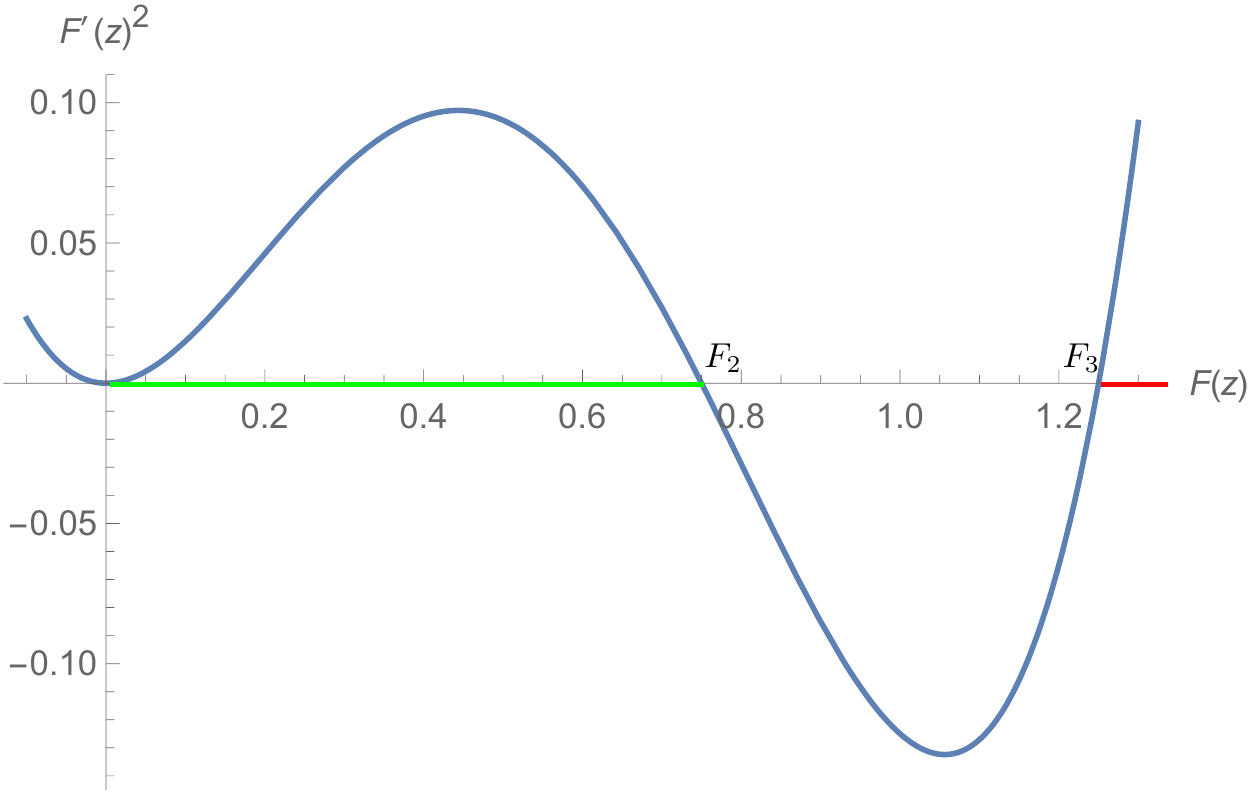}
}
\subfigure[]{
\includegraphics[scale=0.4]{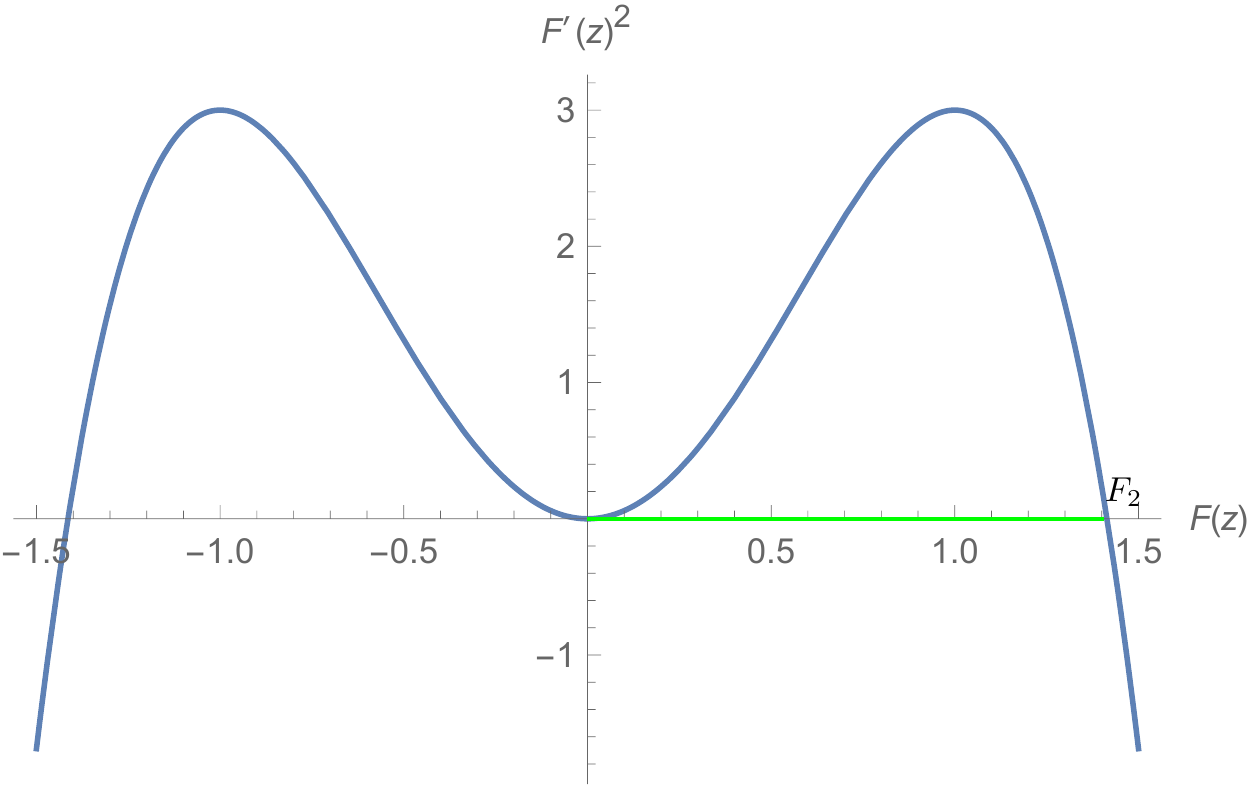}
}
\subfigure[]{
\includegraphics[scale=0.4]{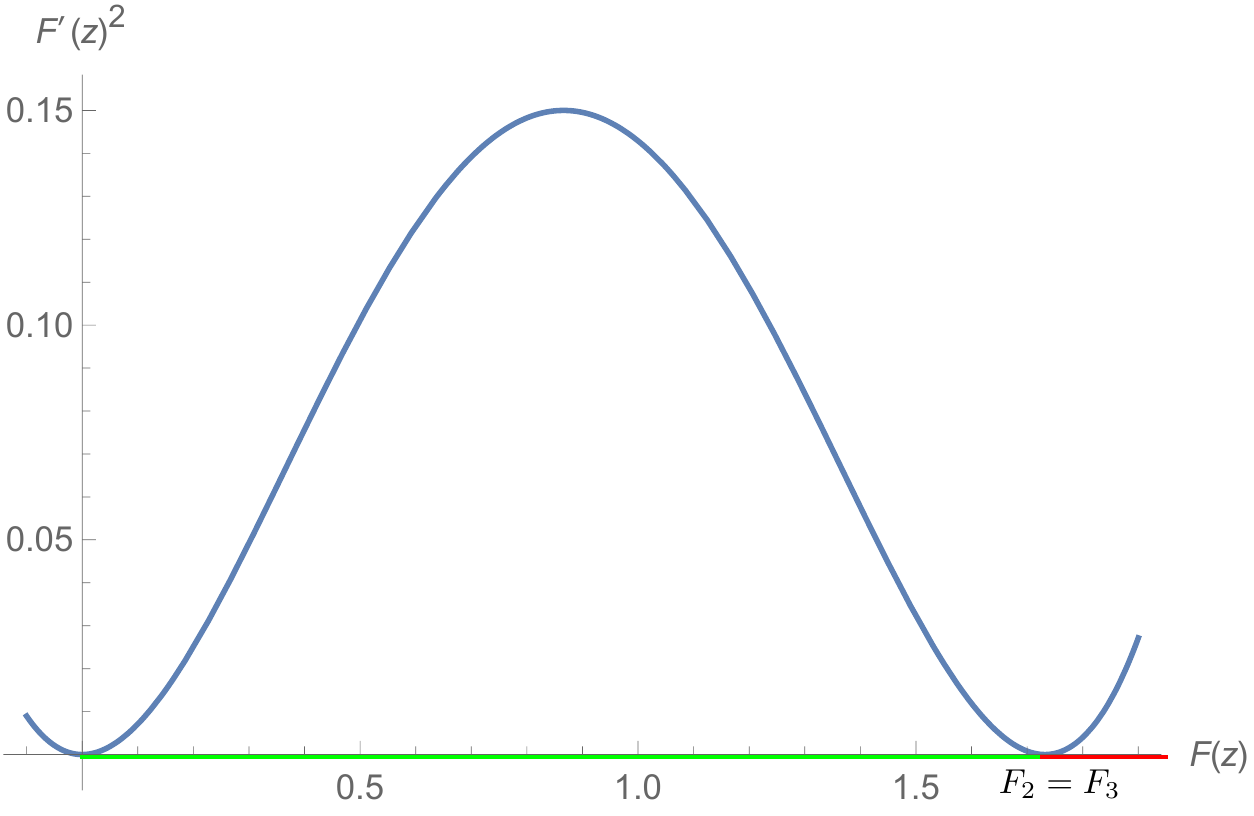}
}
\subfigure[]{
\includegraphics[scale=0.4]{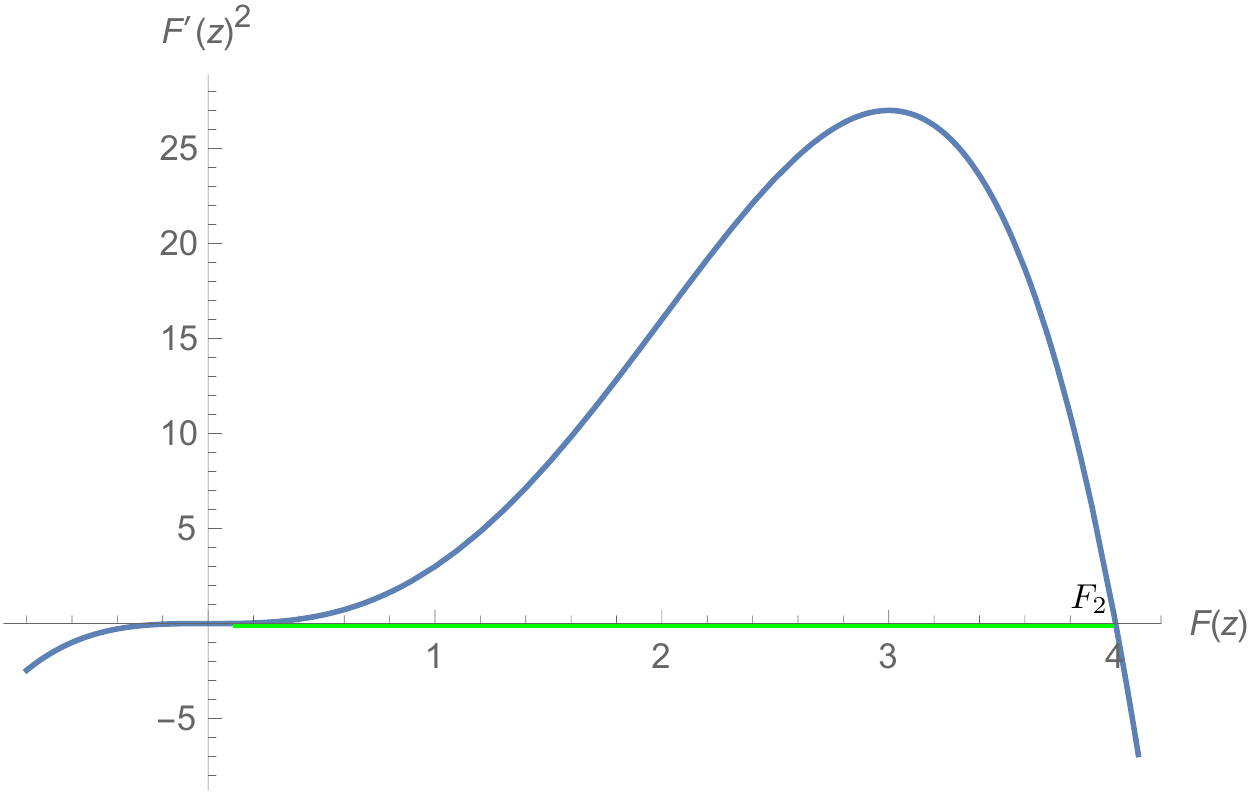}
}
\subfigure[]{
\includegraphics[scale=0.4]{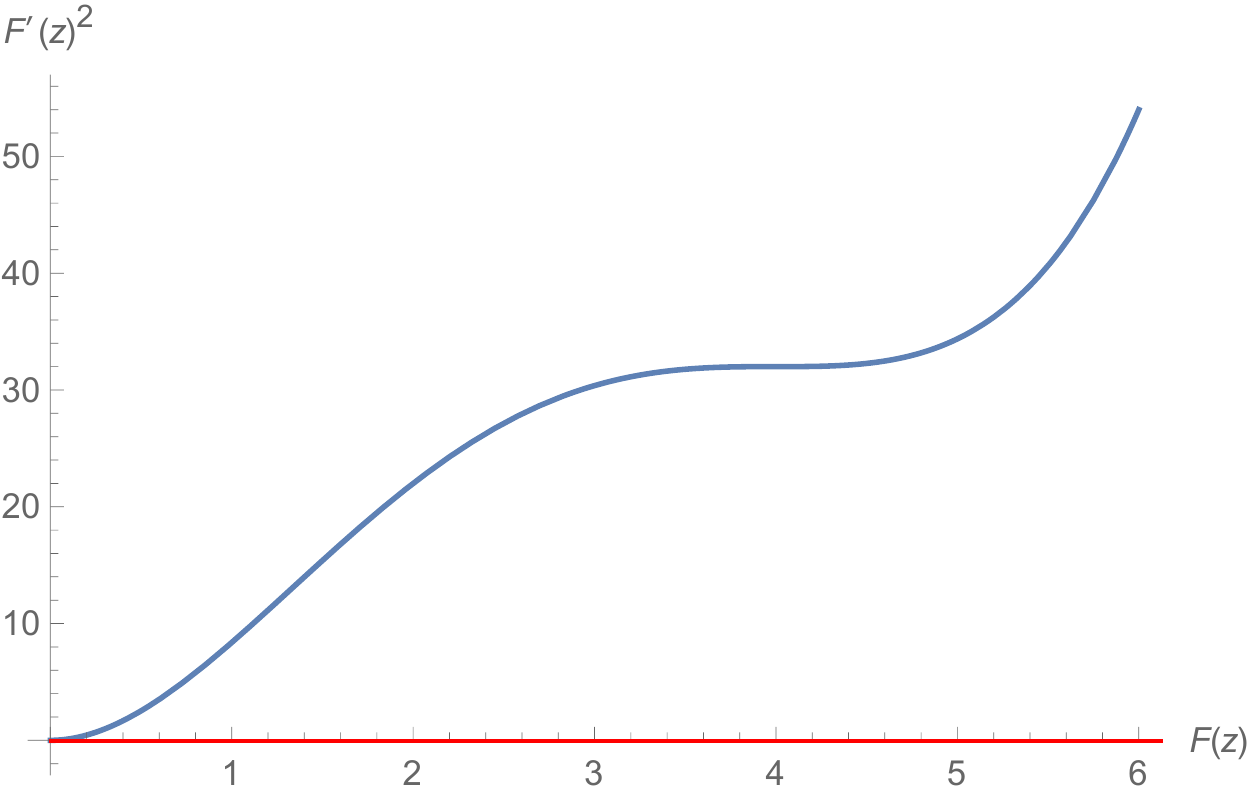}
}
\caption{Phase diagrams (types) $\{(F'(z))^2, F(z)\}$ according to Eq.(37) corresponding to bounded or singular solutions $F(z)$. Green domains: bounded solutions; red domains: singular solutions.
(a): $0<F_3\le F_0\le F_2,\alpha<0,\gamma<0,\beta>0,3\gamma\alpha<2\beta^2$, periodic $F(z)$.
(b): $0<F_0\le F_2,\alpha>0,\gamma>0,\beta<0,3\gamma\alpha<2\beta^2$, pulse-like $F(z)$; $F_0>F_3$, singular. (c): $0<F_0\le F_2,\alpha<0,\gamma>0,\beta \in \mathbf{R}$, pulse-like $F(z)$. (d): $0<F_0\le F_3=F_2,\alpha>0,\gamma>0,\beta<0,3\gamma\alpha=2\beta^2$, kink-like $F(z)$; $F_0>F_2=F_3$, singular. (e): $0<F_0\le F_2,\alpha<0,\gamma=0,\beta>0$, algebraic pulse-like $F(z)$. (f): $2\beta^2<3\gamma\alpha, \alpha>0,\gamma>0$, unbounded $F(z)$ for any $F_0>0$. Comments in the text.}
\end{figure}

It is clear that  the choice of the intensity $F_0>0$ in relation to the zeros of $(F')^2$ ($F_1=0, F_{2,3}=-\frac{2\beta\pm\sqrt{4\beta^2-6\alpha\gamma}}{\alpha}$)
is essential for the singularity behaviour of $F(z)$. $F_0>0$ must be chosen such that $(F')^2_{|_{F=F_0}}\ge 0$. In this manner certain domains are defined 
(labelled green or red in Fig.1), where $(F'(z))^2$ is bounded (in a finite interval) or unbounded (in an infinite interval), respectively.
Characterisation in the phase diagram conditions (PDCs) (a)-(f) in the captions of Fig.1 (periodic, pulse-like, kink-like) is well known in the literature \cite{DSch}. 

The foregoing results can be summarised as follows. Bounded or unbounded solutions $\Re(\psi(x,t))=\sqrt{F(x-ct)}cos(\phi(x-ct)-\omega t)$ of Eq.(2) exist if $\phi'(z)$ and $F(z), z=x-ct,$ are related by $\phi'(z)=d\frac{F'(z)}{F(z)}$ or $\phi'(z)=b_0+b_1F(z)$, if the amplitude $F(z)$ satisfies the ODE $(F'(z))^2=\alpha F^4(z)+4\beta F^3(z)+6\gamma F^2(z)$,
and if the parameters of the QCGLE satisfy the PDCs associated to Fig.1 together with certain constraints (see, e.g., (21),(35)). The parameter range for existence of these particular solutions is the subspace (in parameter space) defined by the PDCs and the constraints.

\section{Examples}

To elucidate the foregoing results, we first consider $\phi'(z)=d\frac{F'(z)}{F(z)}$ with $c=0$. Needless to say, that if all parameters are prescribed, constraints (21) are not satisfied in general. Due to (21) one of the parameters $\{\epsilon, h_j,c_j, j=1,3,5\}$ cannot be prescribed. As a solution of (21), this parameter must be inserted into (20) leading to (lengthy) expressions for $d$, hence for $\omega,\alpha,\beta,\gamma,g_2,g_3$ and, finally, to $F(x)$ and $\phi(x)$ according to (39) and (42), respectively. For simplicity, we assume a particular solution  of (21) w.r.t. $h_3$ and $h_5$. If
\begin{equation}
h_3=\frac{2c_3h_1}{c_1},\quad h_5=\frac{3c_5h_1}{c_1},\quad sign\frac{c_3}{c_1}=\pm 1,
\end{equation}
the constraint (21)
$$
\frac{-3D_3\mp\sqrt{9D_3^2+8D_2^2}}{4D_2}=\frac{-2D_5\pm\sqrt{4D_5^2+3D_4^2}}{2D_4}
$$
(where $\mp$ and $\pm$ corresponds to sign in (44)) is satisfied and  $c$ is real (see Eq.(34)), we obtain
\begin{equation}
d=-\frac{c_1}{2h_1},\quad \omega=-\frac{\epsilon c_1}{h_1}
\end{equation}
\begin{equation}
\alpha=\frac{4c_5h_1^2}{c_1D_1},\quad \beta=\frac{c_3h_1^2}{c_1D_1},\quad \gamma=-\frac{2\epsilon h_1}{3D_1}
\end{equation}
\begin{equation}
g_2=\frac{4\epsilon^2 h_1^2}{3D_1^2},\quad g_3=\frac{8\epsilon^3 h_1^3}{27D_1^3}.
\end{equation}
Subject to (47) the PDC according to Fig.1(a)-(f) must be evaluated. For instance, parameters
\begin{equation}
c_1=-1,\quad c_3=-1,\quad c_5=\frac{1}{8},\quad h_1=-1,\quad h_3=-2,\quad h_5=\frac{3}{8},\quad \epsilon=-1,\quad F_0=4
\end{equation}
are consistent  with the PDC of Fig.1(a). In this case $F(x)$ (see Fig.2) is periodic with period $p$ (see \cite{Abr}, 18.12.30)
\begin{equation}
p=\frac{\pi \sqrt{D_1}}{\sqrt{\epsilon h_1}}.
\end{equation}
A plot of $\Re(\psi(x, t))=\sqrt{F(x)}cos(\phi(x)-\omega t)$ with $\phi(x)$ according to (42) is shown in Fig.3.

The second case $\phi'(z)=b_0+b_1F(z)$ can be exemplified following the line presented before. If
\begin{equation}
c_5=\frac{h_5(3c_1c_3^2-4c_3h_1h_3-c_1h_3^2)}{c_3^2h_1+4c_1c_3h_3-3h_1h_3^2},
\end{equation}
constraint (35) is satisfied. Subject to (51), parameters 
\begin{equation}
c_1=-\frac{3}{4},\quad c_3=1,\quad c_5=-2.5,\quad h_1=1,\quad h_3=-1,\quad h_5=-1,\quad \epsilon=1,\quad F_0=\frac{7}{32}
\end{equation}
are consistent with the PDC of Fig.1(d). Coefficients $\alpha,\beta,\gamma$ and invariants $g_2,g_3$ are given by (37) and (40),
respectively. Two field patterns of $\Re(\psi(x, t))$ are shown in Figs.4(a) and 4(b). We note that the PDC according to Fig.1(d) is the only possible, since $\alpha,\beta,\gamma$ according to (39) imply $3\alpha\gamma=2\beta^2$, independent on the choice of $\{\epsilon, h_j, c_j, j=1,3,5\}$. As mentioned above, the unbounded ("spiky") solution $F(z)$, depicted in Fig.4(b), appears because $F_0$ is greater than the double root $-\frac{3\gamma}{\beta}(\approx 1.73)$ of $(\alpha F^4(z)+4\beta F^3(z)+6\gamma F^2(z)=0$.

As is known \cite{SS2}- \cite{SS4} phase diagram analysis is an effective approach to study existence and parameter dependence of the solutions $F(z)$ of the nonlinear ODE (36). For instance, parameters (49) correspond to a physical (bounded and nonnegative) solution $F(x)$ according to (39) and hence to a physical $\psi(x,t)$. The associated PDC together with the corresponding constraint(s) 
define a subspace and thus a range of parameter variation. The (particular) constraint 
$$
\frac{-3D_3-\sqrt{9D_3^2+8D_2^2}}{4D_2}=\frac{-2D_5+\sqrt{4D_5^2+3D_4^2}}{2D_4}
$$
is satisfied by (45). Thus the range $\{F_0,\epsilon,c_1,c_3,c_5,h_1\}$ is defined by the PDC of Fig.1(a) only. With parameters (see (49)) $c_1=-1,\quad c_3=-1,\quad h_1=-1,\quad h_3=\frac{2c_3h_1}{c_1},\quad h_5=\frac{3c_5h_1}{c_1}$,
evaluation of the PDC leads to the range $\{F_0,\epsilon,c_5\}$ depicted in Fig.5. With parameters  $c_5=\frac{1}{8},\quad h_3=\frac{2c_3h_1}{c_1},\quad h_5=\frac{3c_5h_1}{c_1}\quad \epsilon=-1,\quad F_0=4$,
evaluation of the PDC yields to the range $\{c_1,c_3,h_1\}$ depicted in Fig.6. Further triples of $\{F_0,\epsilon,c_1,c_3,c_5,h_1\}$ can be considered correspondingly. Thus the whole range of parameter variation of (49), associated to solution $\Re(\psi(x,t))$, depicted in Fig.3, can be determined. -- Needless to say, this kind of considerations
also applies to solutions corresponding to Figs.1(b)-(e).

\section{Conclusion}

In conclusion, we presented an approach to obtain closed-form traveling wave solutions of the QCGLE. The central assumptions are restrictions on the dependence between $\phi'(z)$ and $F(z)(= f^2(z))$ according to Eqs.(5) and (6). The solution $F(z)$ is compactly represented by Eq.(39) in terms of Weierstrass elliptic function $\wp(z,g_2,g_3)$ (disregarding the fact that $\wp$ is degenerating due to the vanishing discriminant of $\wp$). As a consequence, the phase function $\phi(z)$ can be represented in closed form analytically (see Eqs.(42) and (44)). The behavior of $F(z)$ is studied by means of a phase diagram approach leading to conditions for "physical" (periodic, pulse-like, kink-like) solutions $F(z)$ as well as to conditions for unbounded ("spiky") solutions $F(z)$. In particular, we obtained the remarkable result that no bounded solution exists if the parameters of the QCGLE satisfy the condition $2\beta^2-3\gamma\alpha<0$, where $\alpha,\beta,\gamma$ are given by (37) or (38) and irrespective of $F_0>0$. -- The phase diagram approach is also suitable to investigate the parameter dependence of solutions (see Section IV). 

Finally, we compare our approach with some other methods for getting solutions of the QCGLE.\\
1) The "simple technique" presented in [8] leads to some results that are consistent with corresponding results above: Assumption (5) in [8] is essentially the same as Eq.(5) above. Results (11) and (12) in [8] are the same as (20) and (19) above, respectively. It seems that (14) in [8] is identical with (21) above. Periodic solutions $F(z)$ are not presented in [8], though they are possible solutions of Eq.(34) in [8], that has the same structure as Eq.(36) above. By using (38) with (10)-(12), it seems that (36) above and (34) in [8] are consistent. Nevertheless, $f(t)$ according to (35) in [8] is not the general solution of Eq.(34) in [8].\\
2) Based on numerical simulations, extreme amplitude solutions of the QCGLE are reported in \cite{Ch}. Whether they are related to the unbounded solutions above (see Fig.4(b)) is an open question.\\
3) The particular relations (3.38 a,b), (3.51 a,b), (3.57 a,b) used in [2] lead to exact solutions  that describe fronts (kinks), pulses, sources and sinks. Obviously, as in [9], periodic solutions are not presented.\\
4) In [9], based on a Laurent expansion ansatz and a particular relation between amplitude and phase function (see Eqs.(12),(14) in [9]), exact solutions are derived (see Eq.(20) in [9]).\\
5) In our estimation, even if we take into account recent publications ([3], and references therein), the most comprehensive general (without particular relations between phase and amplitude) treatment to obtain exact solutions of the QCGLE is presented in [5], [4a], [11, Eqs.(33),(34)], [4b, Eq.(52)], [6], [7, Eq.(50)]. Comparing the solutions in [4b] and [11] with solutions (39), (42) and (43), (44) above, we first note that the number of constraints are different. Solutions (37), (42) are associated to \underline{two} constraints as one of the four possibilities of (21); the particular solutions (33), (34) in [11] are connected with 3 and 5 constraints, respectively. As mentioned above, the general solution (52) in [4b] exists subject to 5 constraints. With respect to the formal representation of amplitude and phase, we secondly note that results in [4b] and [11] are simpler than our results above. The same holds for Eq.(50) in [7]. However it seems that -- in the general treatment -- the question of defining conditions for physical solutions is not addressed. This is not a trivial problem (if, for example, solution (52) in [4b] is considered, meromorphic $M$ must be non-negative and $\phi'$ must be real. These requirements must be discussed subject to the (well known) analytical properties of $\wp^2(\xi)$ and $\wp(2\xi)$ and subject to 5 constraints). -- Obviously, the phase diagram approach is suitable not only for studying the parameter dependence of solutions outlined above, but also for specifying the parameters such that solutions (39), (42), (43), (44) and hence $\Re(\psi(x,t))$ are physical.\\
6) For the cubic Ginzburg-Landau equation the non-existence of elliptic traveling wave solutions has been proved \cite{H}.
So it is obvious to consider our result above if $c_5=h_5=0$. With $\phi'(z)=d\frac{F"(z)}{F(z)}$ we obtained $c=0$ necessarily,
consistent with the proposition in \cite{H}. According to (20), (21) it follows $d=0$ and hence $\phi(z)=const$. For case $\phi'(z)=b_0+b_1F(z)$, we assumed 
$b_0=\frac{cc_1}{2D_1}$ for simplicity, leading to $b_1=0$ (see (32), (33)), and $b_0\to \infty$, in contradiction to bounded $\phi'(z)$.

Summing up, for both cases (5) and (6), our results are consistent with \cite{H}. Without assuming $b_0=\frac{cc_1}{2D_1}$ the question of consistency 
is open. -- With respect to case (6), we conclude, that the non-existence claim in \cite{H} is not valid in the cubic-quintic case. 

Directions in which further investigations can go should be indicated. First, it would be interesting to find more relations than (5) and (6) in order to increase the solution set of the QCGLE. Secondly, it would be important to find generalisations for ansatz (5) as well as for (6) by, for example, including a further parameter. Thirdly, if (5) and (6) are modified by "small corrections", the solutions presented above maybe taken as start solutions (different from solutions of the NLSE) for a perturbation approach. Fourthly, a stability analysis of the solutions found with respect to the parameters of the QCGLE as well as with respect to "small" perturbation of $f$ seems possible. Finally,
we note that all traveling wave solutions of Eq. (2) must satisfy the system (3) - (4). - It is an interesting task to check consistency
of this fact with solutions obtained by the general procedure outlined in [4] - [7], [9], [10] - [12].

\begin{figure}
\centering
\includegraphics[scale=0.6]{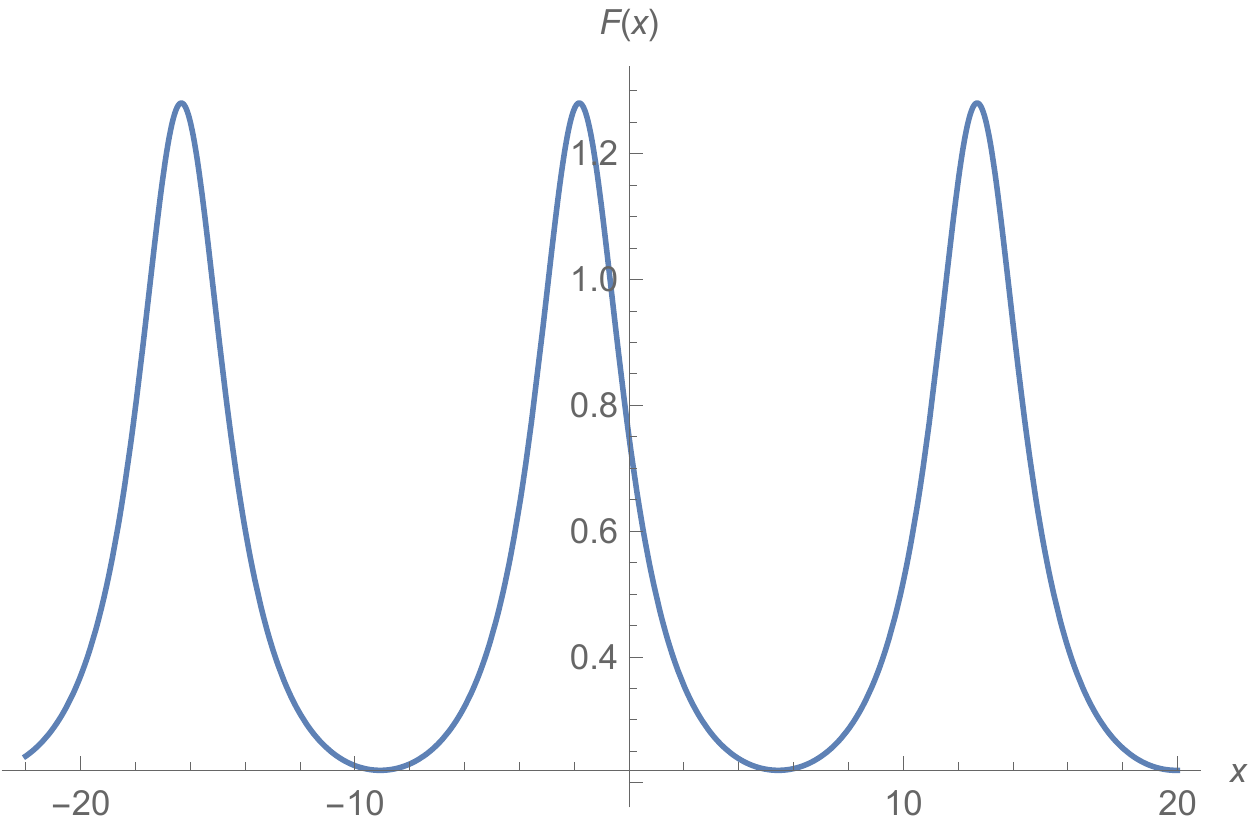}
\caption{Intensity $F(x)$ according to Eq.(37) and parameters (48).}
\end{figure}

\begin{figure}
\centering
\includegraphics[scale=0.6]{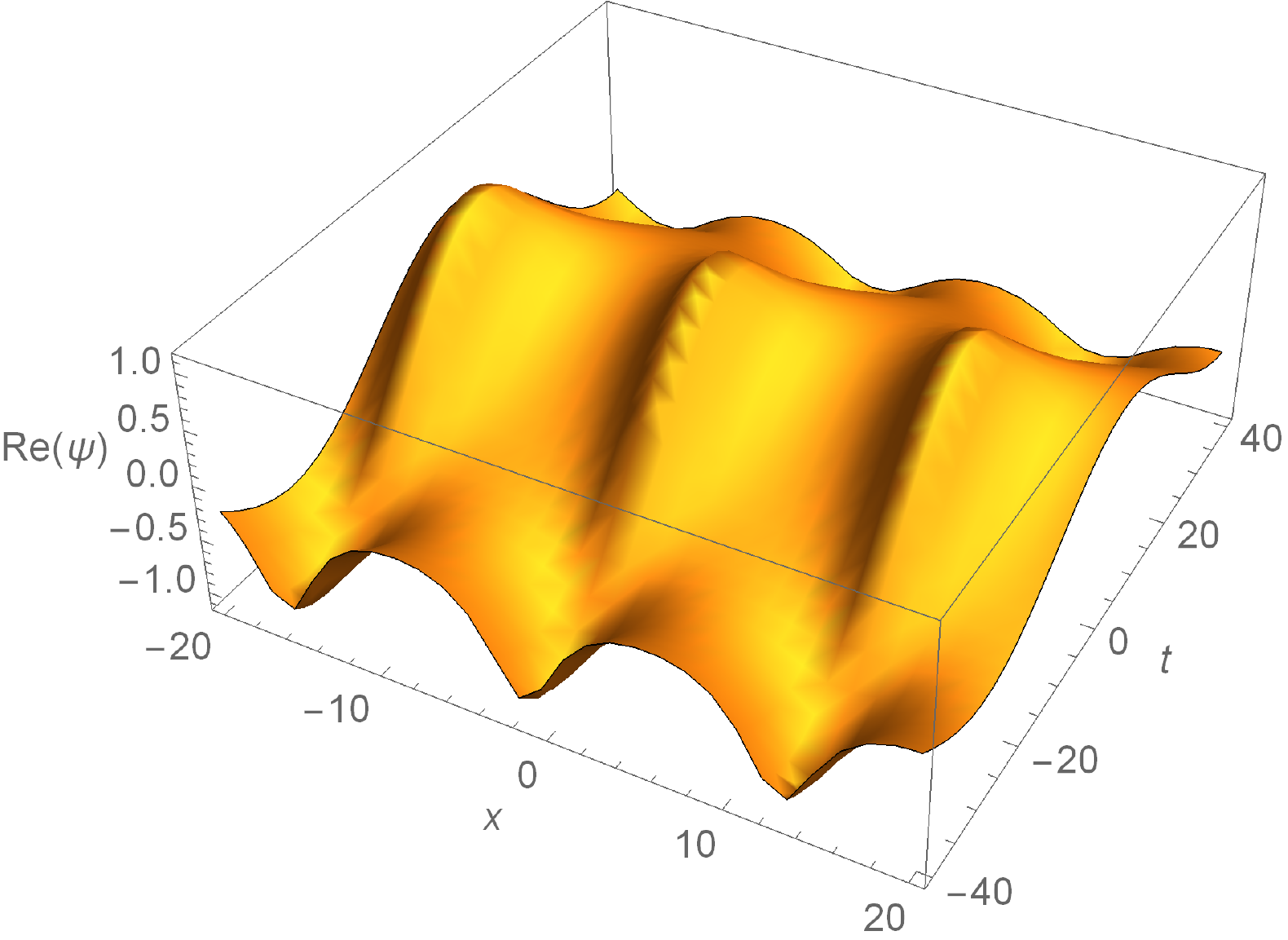}
\caption{Field pattern according to Eq.(1) and parameters (48).}
\end{figure}

\begin{figure}
\subfigure[]{
\includegraphics[scale=0.5]{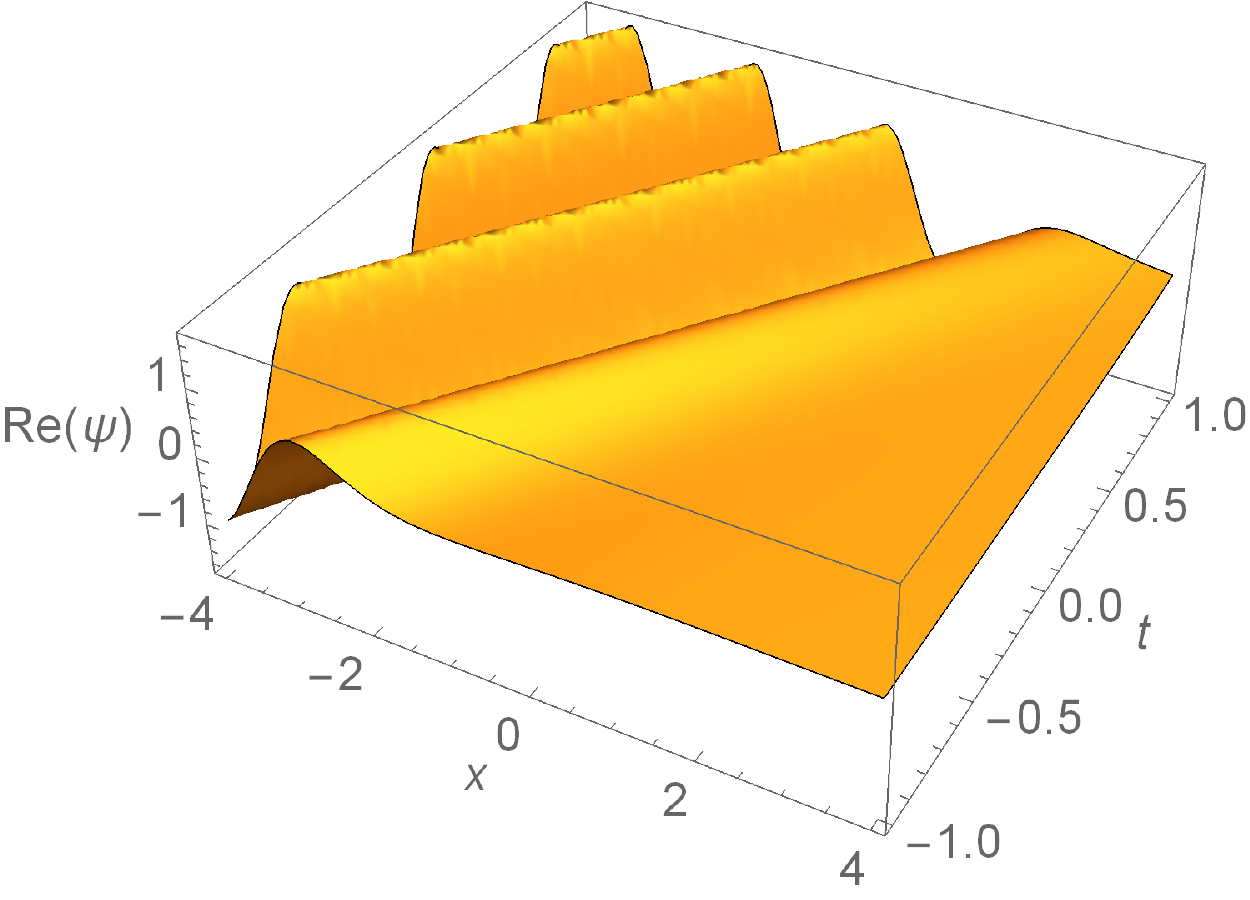}
}
\subfigure[]{
\includegraphics[scale=0.6]{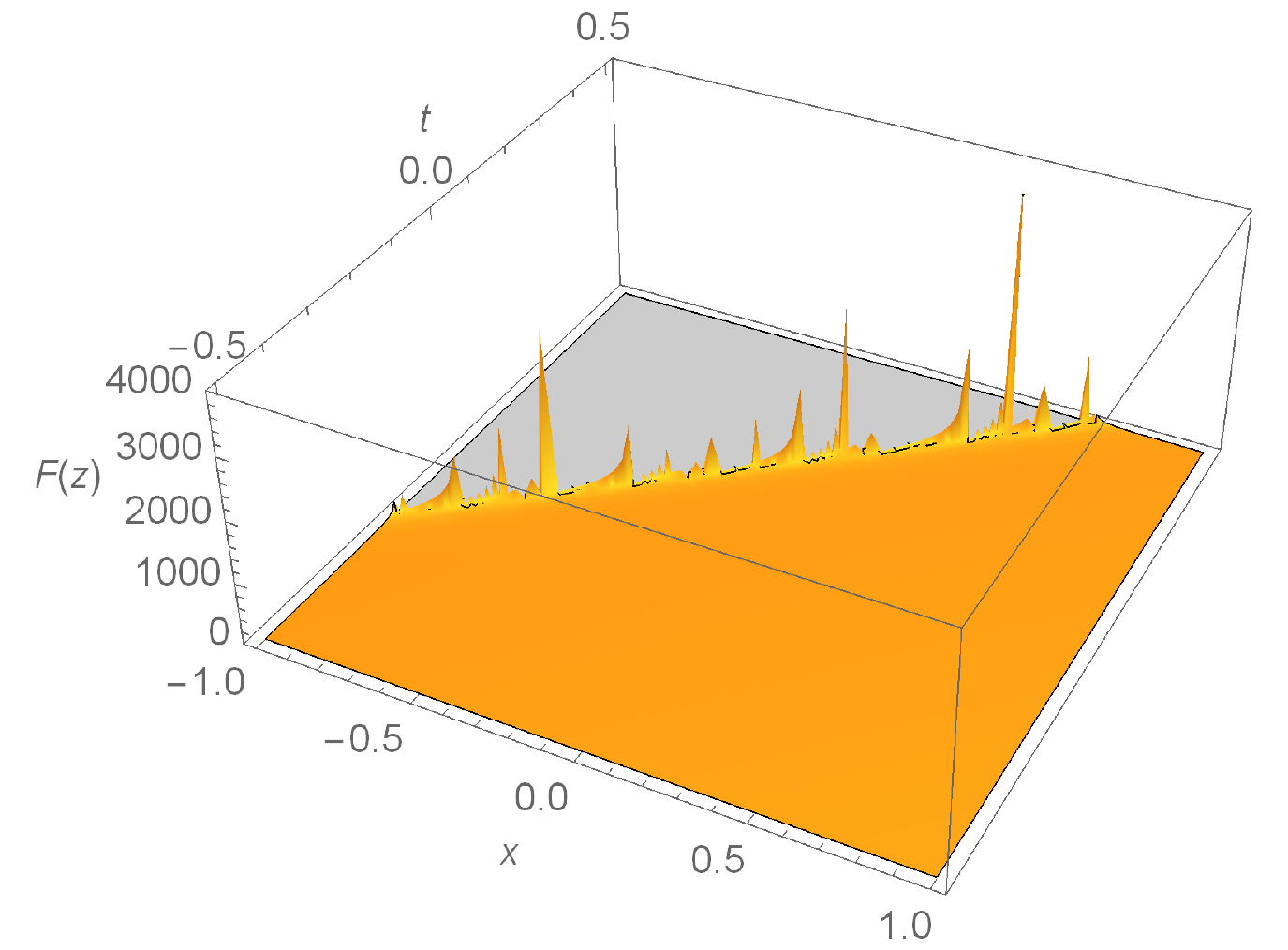}
}
\caption{Field patterns according to Eq.(1) and parameters (51); (a): $F_0=\frac{7}{32}$. (b): $F_0=6$. Comments in the text.}
\end{figure}

\begin{figure}
\centering
\includegraphics[scale=0.6]{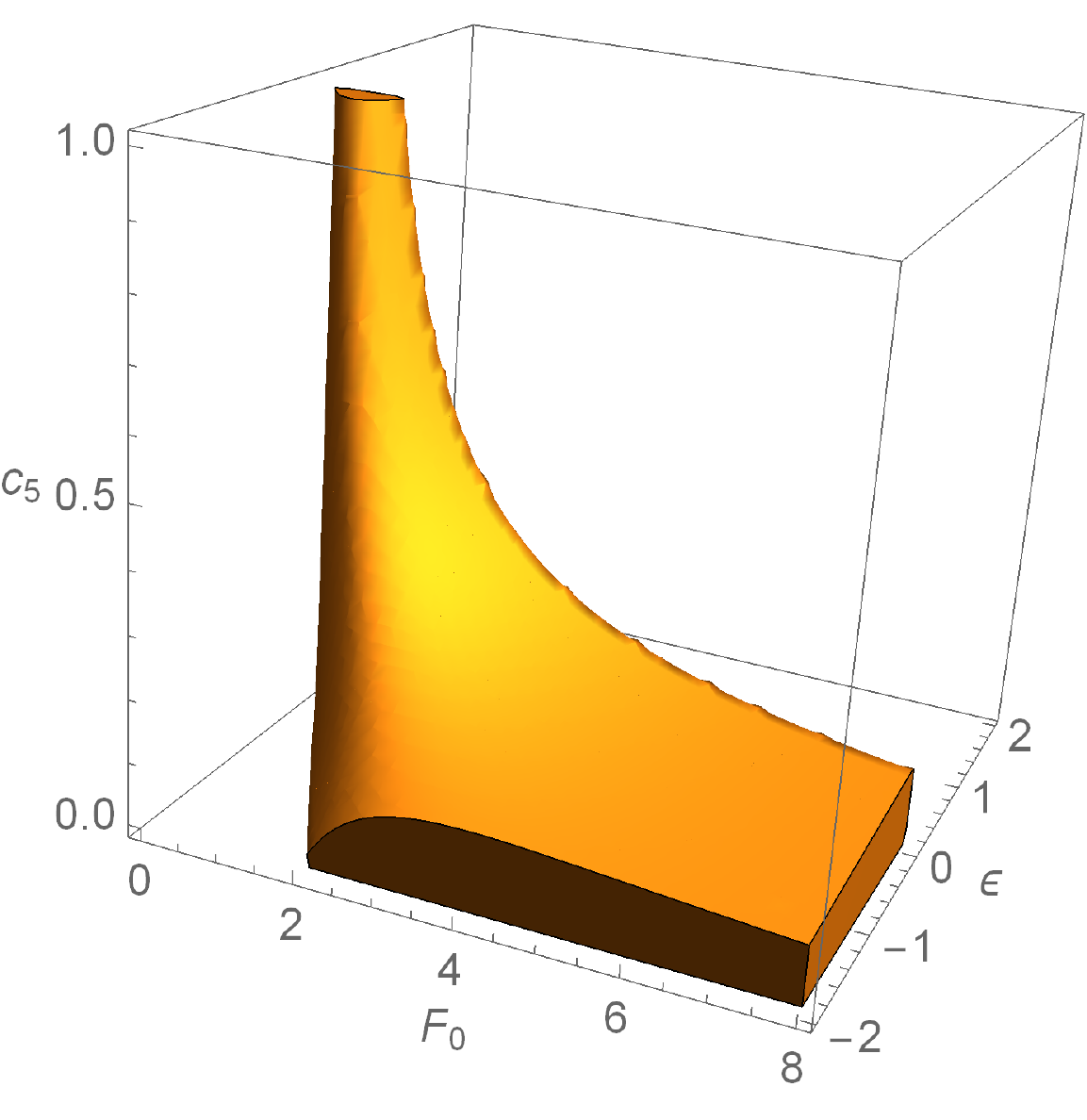}
\caption{Parameter region for allowed $\{F_0,\epsilon,c_5\}$ according to phase diagram Fig.1(a) and parameters: $c_1=-1,c_3=-1, h_1=-1, h_3=-2, h_5=0.375$.}
\end{figure}

\begin{figure}
\centering
\includegraphics[scale=0.6]{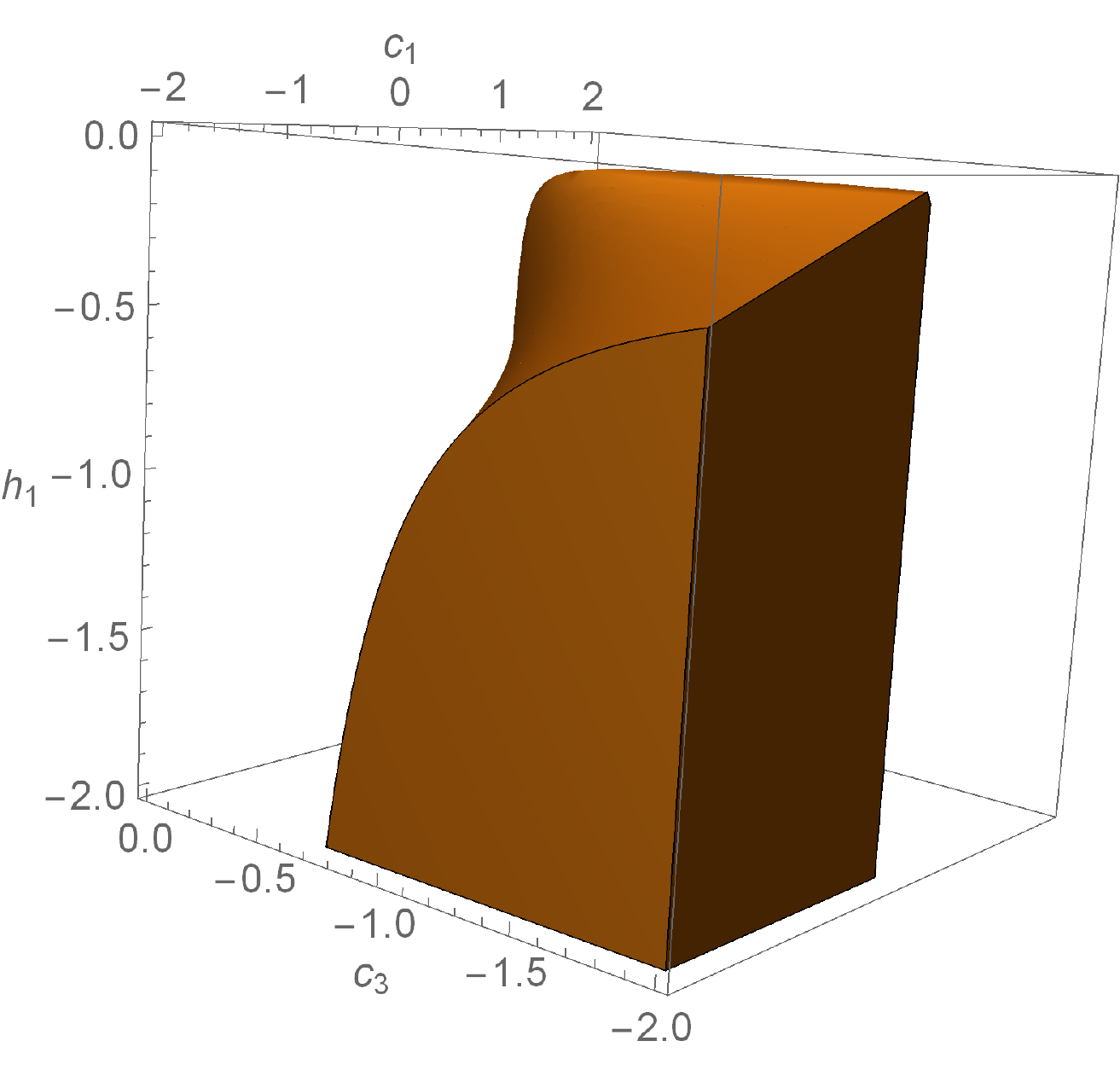}
\caption{Parameter region for allowed $\{c_1,c_3,h_1\}$ according to phase diagram Fig.1(a) and parameters: $c_5=\frac{1}{8}, h_3=-2, h_5=0.375, \epsilon=-1, F_0=4$.}
\end{figure}

\end{document}